\documentclass[twocolumn,showpacs,preprintnumbers,amsmath,amssymb]{revtex4}
\usepackage{graphicx}
\usepackage{dcolumn}
\usepackage{bm}
\usepackage{here}

\begin{document}

\title{Analyticity of Quantum States in One-Dimensional Tight-Binding Model}
\author{Hiroaki S. Yamada}
\email{hyamada[at]uranus.dti.ne.jp}
\affiliation{Yamada Physics Research Laboratory,
Aoyama 5-7-14-205, Niigata 950-2002, Japan}
\author{Kensuke S. Ikeda}
\email{ahoo[at]ike-dyn.ritsumei.ac.jp}
\affiliation{Department of Physics, Ritsumeikan University
Noji-higashi 1-1-1, Kusatsu 525, Japan}

\date{\today}
\begin{abstract}
Analytical complexity of quantum wavefunction whose argument
is extended into the complex plane provides an important
information about the potentiality of
manifesting complex quantum dynamics such
as time-irreversibility, dissipation and so on.
We examine Pade approximation and some complementary methods
to investigate the complex-analytical properties of some
quantum states such as impurity states, Anderson-localized states
and localized states of Harper model.
The impurity states can be characterized by simple poles of
the Pade approximation, and the localized states of
Anderson model and Harper model
can be characterized by an accumulation of
poles and zeros of the Pade approximated function
along a critical border, which implies
a natural boundary (NB). 
A complementary method based on shifting
the expansion-center is used to confirm the existence of the
NB numerically, and it is strongly suggested that the
both Anderson-localized state and localized states of Harper model
have NBs in the complex extension.
Moreover, we discuss an interesting relationship between
our research and the natural boundary problem of the potential
function whose close connection
to the localization problem was
discovered quite recently by some mathematicians.
In addition, we examine the usefulness of the Pade approximation
for numerically predicting
the existence of NB by means of two typical examples,
lacunary power series and random power series.
\end{abstract}


\maketitle


\def\ni{\noindent}
\def\nn{\nonumber}
\def\bH{\begin{Huge}}
\def\eH{\end{Huge}}
\def\bL{\begin{Large}}
\def\eL{\end{Large}}
\def\bl{\begin{large}}
\def\el{\end{large}}
\def\beq{\begin{eqnarray}}
\def\eeq{\end{eqnarray}}

\def\eps{\epsilon}
\def\th{\theta}
\def\del{\delta}
\def\omg{\omega}

\def\e{{\rm e}}
\def\exp{{\rm exp}}
\def\arg{{\rm arg}}
\def\Im{{\rm Im}}
\def\Re{{\rm Re}}

\def\sup{\supset}
\def\sub{\subset}
\def\a{\cap}
\def\u{\cup}
\def\bks{\backslash}

\def\ovl{\overline}
\def\unl{\underline}

\def\rar{\rightarrow}
\def\Rar{\Rightarrow}
\def\lar{\leftarrow}
\def\Lar{\Leftarrow}
\def\bar{\leftrightarrow}
\def\Bar{\Leftrightarrow}

\def\pr{\partial}

\def\Bstar{\bL $\star$ \eL}

\def\etath{\eta_{th}}
\def\irrev{{\mathcal R}}
\def\e{{\rm e}}
\def\noise{n}
\def\hatp{\hat{p}}
\def\hatq{\hat{q}}
\def\hatU{\hat{U}}

\def\iset{\mathcal{I}}
\def\fset{\mathcal{F}}
\def\pr{\partial}
\def\traj{\ell}
\def\eps{\epsilon}

\section{Introduction}
Poincare has proved that 
almost all the classical systems 
are nonintegrable and the phase space of the nonintegrable systems 
contain chaotic invariant set almost everywhere,
although their measure might not be appreciably large as proved by 
KAM theory \cite{gutzwiller91}.
The chaotic sets play a crucial role for the formation of the local and
global paths which connect every physically significant portions 
in the phase space and realizes statistical ensemble as is 
typically exemplified by the microcanonical ensemble. 
On the chaotic sets 
it is very hard to reverse the evolution of  classical 
chaotic dynamics in time because
of the exponential orbital instability.
The instability results in irreversible loss of
phase space information and  may cause the onset of dissipative phenomena. 
Accordingly, the role of the chaotic instability is particularly 
important as an origin of irreversibility 
in the classical systems with  
small number of degrees of freedom.

On the other hand, in closed quantum systems, 
the strength of dynamical instability is 
much weakened \cite{yamada96,yamada02};
the exponential instability of classical chaos is not allowed 
because of the discreteness of energy levels, and, more basically, 
because of the uncertainty principle that do not allow
the emergency of arbitrary small scale phase space structure
resulting from the chaotic instability. 
It causes the localization effect
which severely suppresses 
the time-irreversible transportation process such as chaotic diffusion
\cite{casati96}.
Indeed, the localization effect was originally discovered for
the low-dimensional disordered quantum system and 
 was considered as a strong manifestation of the robustness of 
quantum coherence \cite{anderson58,abrahams10}.
It is ordinarily emphasized that the quantum dynamics is coherent
and exhibits the typical interference phenomena which seems to be often
classically paradoxical \cite{aharanov59,bell04}.
The coupling with macroscopic degrees of freedom surely makes 
the quantum system irreversible and incoherent \cite{caldeira85},
however, if the system is isolated from the macroscopic degrees of freedoms, 
the coherent and interference nature seems to be a stable 
characteristics of quantum systems, 
which is strongly desired in the application to the quantum 
communication \cite{nielsen00}.

In spite of the handicaps of quantum dynamics for actualizing
the time-irreversibility, quantum systems exhibits surprising ability
of restoring the time-irreversibile natures under some weak conditions. 
In low-dimensional quantum disordered systems and quantum 
chaos systems the localization effect mentioned above is destroyed by 
the couplings with other degree of freedom 
\cite{casati79,adachi88,miller99,yamada02} and/or increments of spatial 
dimensionality \cite{goda99,kuzovkov04}. 
Indeed, extensive investigations of the disordered quantum systems
and quantum chaos systems lead us to a completely opposite view about 
the quantum dynamics \cite{peres84,shiokawa95,gorin06,jacquod09,yamada12a,yamada12b}: 
it is naturally unstable and exhibits 
time-irreversibility even if its number of degrees of freedom is small
and do not couple with macroscopic degrees of freedom.
Moreover, the systems often undergoes a phase transition to spontaneously 
recover an apparently irreversible nature if the
coupling strength with 
 the degrees of freedom exceeds a critical value
decided by Planck constant as we demonstrated 
in Refs.\cite{ikeda93,kolovsky94}.

The attempts we would like to present here is to discuss
{\it what mathematical structure of the wavefunction
is responsible for the irreversibility in quantum
dynamics.}
We take the quantum wavefunctions
before the transition to the apparently time-irreversible 
states, and investigate its analytical structure
by extending the argument of wavefunction, which is 
usually a real variable, into the complex plane.
In the concrete, we examine singularity of various localized states of
the one-dimensional tight binding models.
This is because 
 symptom of anomalous singularity of these quantum states
 may be developed in the extended complex plane
although these quantum states 
are analytically smooth and shows no anomalous features
on the real space due to the normal convergence.

We present some explicit results for localized quantum states 
of Harper model
\cite{harper55,last95,jitomirskaya95,wilkinson94,hiramoto88,
aubry80,ostlund84,ketoja95,satija99,Jazaeri01}
and also for the 
localized quantum states, namely, impurity states
and Anderson localized states of disordered one-dimensional tight binding
models.\cite{ishii73,lifshiz88,stollmann01}.
As far as we know, this is the first attempt
to investigate the complex analytical characteristics
of wavefunction in a closed quantum system.

It should be compared with the
multifractal analysis to investigate morphological
complexity of the wavefunctions.
Multifractal analysis as well as scaling analyses of
wavefunction have been already used
to characterize the statistical fluctuation of the wavefunction
in the some tight-binding models
\cite{aoki83,soukoulis84,pietronero87,siebesma87,schreiber90,berry96,wojcik00}.
However,the multifractal analysis limit themselves
to the analysis of the singularity on the real coordinates.
On the other hand, we are concerned with singularities of the 
wavefunction in the complex plane. It may be even said that 
the real space singularity is an occasional realization 
of more general complex-space singularities on the real plane. 
In a previous paper, we regarded the Anderson localized state
as a ``potentially irreversible'' quantum state \cite{yamada02}
although it evidently lacks the time-irreversible nature. 
We are thinking that the singularities in the complex space 
contains the essential information about the ``potential time-irreversibility''
 of the quantum systems. 
In particular, we are interested in the accumulation of 
singularities which is called {\it natural boundary} (NB).
It is not possible to analytically continue a function given
in a certain domain when singularity points accumulate along
the border of domain.
Then we can say that the function has a NB
beyond which there is no way of extending the function
analytically outside of the domain.
This is the very motivation why we are interested in the singularities
in the complexified observables.

Unfortunately we know no example of time-irreversible system
(with small number of degrees of freedom) which allows analytical
investigation, and so we have to develop numerical method elucidating 
the singularity properties of quantum states. For this
purpose, we adapt the following two numerical methods; Pade approximation, 
and expansion-center-shift method. The Pade approximation has been used
to detect the singularity of functions
by the distribution of poles and zeros of the
approximated rational functions \cite{baker70,baker75,baker96}.
In particular, we are interested in the role of
numerical Pade approximation
for functions with the NB and/or unknown singularity
\cite{yamada13a,yamada13b}.

In Sect.\ref{sect:pade}, we give a brief explanation
for the Pade approximation, and apply it to 
the lacunary and random power series that are known
to have the NB, and we confirm the numerical applicability of 
the Pade approximation for the 
 functions with a NB.

In Sect.\ref{sect:direct-nb}, we explain the expansion-center-shift 
method which decides the positions of the singularities by shifting 
the expansion center and evaluating the convergence.
It can be used for confirming the properties 
of singularities predicted by the Pade approximation from a different
view point. 


In Sect.\ref{sect:pade-quantum_nb}, we present the results
of the Pade approximation
applied to the quantum states of tight-binding models.
Impurity state and Anderson-localized states of disordered tight-binding 
models and localized states of Harper model are examined.
It is suggested that the impurity states have only simple poles
but Anderson-localized states and localized states of Harper model
have the NB in the complex plane.

In Sect.\ref{sect:direct-quantum-nb},
The expansion-center-shift is applied to the three cases examined
in Sect.\ref{sect:pade-quantum_nb}, and 
the results claimed there are confirmed.

In Sect.\ref{sect:quantum-singularity},
we discuss an interesting relationship between
the singularity structures associated with the potential 
and the localization problem
in the one-dimensional tight-binding model
developed by Breuer and Simon \cite{breuer11,costin}.

In the last section, we summarize the results and
give a discussion about the physical meaning
of the results and the connection
with occurrence of irreversibility
in the quantum systems.

In appendices, some additional data including a result
for a dynamically localized state in the Harper model are given.
And some mathematical theorems for
lacunary power series and random power series which are useful
in reading the main text
are also given \cite{ahlfors66,korner93,remmert10}.

\section{Pade Approximation and Natural Boundary:An overview}
\label{sect:pade}
In this section we briefly introduce Pade approximation
as a numerical method to detect the singularity of some functions
with a NB.

\subsection{Pade Approximation}
For a given function $f(z)$ a truncated Taylor expansion about zero
of order $N$ is given as,
\begin{eqnarray}
f^{[N]}(z) = \sum_{n=0}^{N} c_n z^n,
\end{eqnarray}
where ${c_n}$ denotes the coefficients of the expansion.
Consider the rational function defined as a ratio of two
polynomials,
\beq
\nn && f^{[L|M]}(z) \equiv \frac{P_L(z)}{Q_M(z)},
\eeq
with
\beq
\nn && P_L(z) = a_0+a_1z+a_2z^2+....a_Lz^L, \\
&& Q_M(z) = 1+b_1z+b_2z^2+.......b_Mz^M.
\eeq
Pade approximation is to approximate $f^{[N]}(z)$
by $f^{[L|M]}(z)$ up to the highest possible order.
A unique approximation exists for any choice
of $M$ and $L$ such that $N=L+M$
within the order $O(z^N)$ as,
\beq
f(z) - f^{[L|M]}(z) = O(z^{L+M+1}),
\eeq
which is sometimes called $[L|M]$ Pade approximation.

Mathematically, Pade approximation is used to
estimate analyticity of functions.
Indeed, Pade approximations are usually superior
to Taylor expansions when the
functions contain poles because of the use of
rational function.
The Pade approximation
is often significant
even beyond the radius of convergence.
Although the convergence of Pade approximation has
not been clarified yet for general functions 
except for some class of functions with simple singularities,
Pade approximation has been applied to investigate
singularities with even more complex structures.
For example, the convergence properties associated with
the conjugate function of KAM invariant tori in
nearly integrable Hamiltonian maps was extensively studied,
and the natural boundary, which is a border of analyticity
on which singularities are accumulated can be captured well by
the Pade approximation
\cite{berretti90,berretti92,falcolini92,llave94b,berretti95,berretti01}.

In particular, note that the $[M|M]$ Pade approximation or
$[M|M+1]$ Pade approximation is equivalent to a truncated
continued fraction \cite{brezinski91}.
Hereafter, we use the diagonal Pade approximation, $L=M$,
of order $M \leq 65$ to estimate the singularity of the functions
because of the good convergence and limitation due to
the round-off errors and other source of errors.
It is not mathematically clear that how poles of Pade approximated
function describe various types of singularities such as 
 brunch cut and natural boundary
although the convergence property of diagonal Pade approximation
established only for meromorphic and single valued functions
\cite{nuttall70,pommerenke73,stahl98}.
The singularity of the function $f(z)$ is approximated by
the $M$ poles of $P_M(z)$ (and also the $M$ zeros of $Q_M(z)$)
when the $[M|M]$ Pade approximation is used for the function.

In the present paper, we are interested in the NB
of quantum wavefunction,
we focus our attention to the Pade approximation of NB.
The numerical examples applying Pade approximation
to some functions with various types of
singularity will be summarized in a separated paper.

As will be discussed in the present paper, NB is by no means a
mathematical object but a physical object which originates a serious
effect on complex quantum physics. Quite recently, it is shown
that the breakup of KAM tori by NB results in a drastic effect
on the tunnelling process in nearly integrable systems \cite{shudo12}.
Moreover, recently,
it has been strongly suggested that the susceptibility of
the two dimensional Ising model
has a NB in the complex temperature plane
associated with the singularity of
the partition function \cite{guttmann96,nickel99,orrick01,chan11}.

\subsection{Application of Pade approximation to functions with
a NB}
\label{sect:nb1}
In this section we examine the applicability of Pade approximation
to investigating the analyticity of some well-known
test functions with a NB on $|z|=1$. 
This will provide a preliminary information about what
occurs in the Pade approximation of the NB. 
Then, we use the angular-representation $f_r(\theta)$ of the functions
with fixing the polar component $r$ appropriately:
\beq
f_r(\theta) = f(z=re^{i\theta})=\sum_{n=0}^{\infty} c_n (re^{i\theta})^n.
\eeq
Note that the modulus $r$ works as a convergence factor of the series
because the series well converges for $r<1$.
Typically we take $r=1$ on the unit circle or $r=0.98$ inside the circle
in the following numerical calculations.

\subsubsection{Example 1: Lacunary series} 
As the first example, we would like to apply Pade approximation
to the following lacunary series
\beq
f_{Fib}(z) = \sum_{n=0}^{\infty} z^{F_n},
\eeq
where $F_n$ is $n$th Fibonacci number.
This function also has natural boundary on $|z|=1$.
For this particular example, one can show that the
Pade approximated function has an exact form given as
\beq
f^{[F_N]}_{Fib}(z) & \sim & f^{[\frac{F_N}{2}|\frac{F_N}{2}]}_{Fib}(z) \\
& = & \frac{A^{F_N}_{Fib}(z)} { 1+ z^{F_{N-4}}-z^{F_{N-2}} }.
\eeq
The explicit form of the numerator $A^{F_N}_{Fib}(z)$ is given as,
\beq
A^{F_N}_{Fib}(z) &=& S_{N-4}(z) \\ \nn
&+& [S_{N-8}(z)+z](f_{N-4}(z)-f_{N-2}(z)) \\ \nn
&+& [2f_{N-3}(z)+2f_{N-2}(z)+f_{N-3}(z)f_{N-6}(z)],
\eeq
where $S_{L}(z)=\sum_{k=0}^{L} f_{k}(z)$, $f_k(z)=z^{F_k}$.
Here we set
$F_{-1}=F_{-2}=....=0$.
Accordingly, the poles of the $[\frac{F_N}{2}|\frac{F_N}{2}]$ Pade
approximation are given by the roots of a lacunary polynomial,
\beq
1+ z^{F_{N-4}}-z^{F_{N-2}} =0.
\label{eq:poly-fibo}
\eeq

\begin{figure}[htbp]
\begin{center}
\includegraphics[width=8cm]{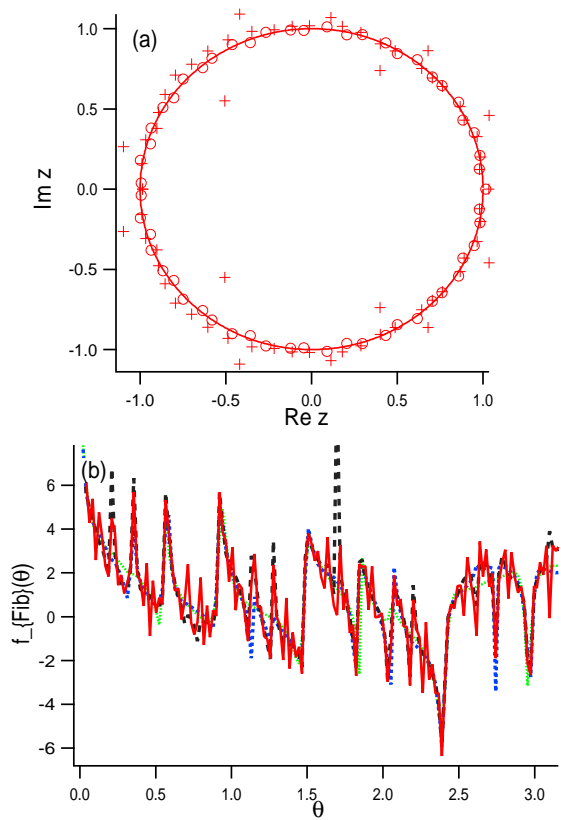}
\caption{
\label{fig:fig4}
(Color online)
(a) Distribution of poles($\bigcirc$) and zeros($+$)
of the $[56|56]$ Pade approximated function
for test function $f_{Fib}(z)$ with a NB on $|z|=1$.
The unit circle is drawn to guide the eye.
(b)The Pade approximated function $f_{Fib,r}^{[56|56]}(\theta)$,
and exact function $f_{Fib,r}(\theta)$
in $\theta-$representation with $r=1.0$.
}
\end{center}
\end{figure}

In Fig.\ref{fig:fig4} the numerical result of
Pade approximation for $f_{Fib,r}(\theta)$ is shown.
The poles and zeros are plotted for the $[56|56]$ Pade approximated
function in Fig.\ref{fig:fig4}(a). 
The poles and zeros accumulate around $|z|=1$
with the increase of $M$.
Some poles coincide numerically with zeros at
very nice precision, and so the poles are cancelled by the zero,
then the poles are ghost poles. The paired pole and
zero is called ``ghost pair''. So the distribution of
unpaired poles are accumulated around $|z|=1$, and well
mimics the NB.

In Fig.\ref{fig:fig4}(b), we express the original function
$f_{Fib,r}(\theta)$
with the $[56|56]$ Pade approximated function, which indicate that the
latter
mimics quantitatively well the former. The singular structure of of original
equation close to the NB is well reproduced.

It can be expressed that 
the complex zeros of the polynomial (\ref{eq:poly-fibo}) cluster near
unit circle $|z|=1$ and the zeros distribute uniformly on the circle
as $F_N \to \infty$ by Erdos-Turan type theorem
in appendix \ref{app:theorems}
\cite{erdos50,amoroso96,odlyzko93,simon04a,simon04b,simon10}.
Furthermore, we give other exact Pade approximated function
for the other lacunary series in appendix \ref{app:lacunary}.

\subsubsection{Example 2: Random power series} 
Let us examine the random power series as the second example:
\beq
f_{noise}(z) &=& \sum_{n=0}^{\infty} \epsilon r_n z^{n}.
\label{eq:noisy-function}
\eeq
Here the random sequence $\{ r_n \}$ is
independently and identically distributed (i.i.d.) random variable
which take the value within $r_n \in [0,1]$, and
$\epsilon$ is the strength of the randomness.
It is shown that in general the noisy function has
a NB on the unit circle $|z|=1$ with probability one.
(See Gilewicz's theorem in appendix \ref{app:theorems}.)

Figure \ref{fig:fig5} shows distribution of poles and zeros of
the $[50|50]$ Pade approximated function for $f_{noise}(z)$.
The poles inside the unit circle $|z|=1$
are cancelled with
zeros at a very nice precision, forming the ghost pairs.
On the other hand, the poles and zeros pairs accumulated
around $|z|=1$ form close pairs but are not exactly cancelled.
Such close pairs are sometimes called as Froissart doublets.
They approximate well the NB of the random power series
on $|z|=1$.

\begin{figure}[htbp]
\begin{center}
\includegraphics[width=6.0cm]{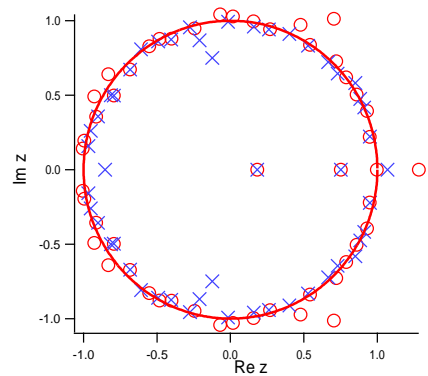}
\caption{
\label{fig:fig5}(Color online)
Distribution of poles ($\bigcirc$) and zeros ($\times$)
of Pade approximated function $f_{noise}^{[50|50]}(z)$
for a random power series $f_{noise}(z)$
with the noise strength $\epsilon=1$.
The unit circle is drawn to guide the eye.
}
\end{center}
\end{figure}


In this case, it has been often noticed that
the complex zeros of the random polynomial also distribute uniformly on
the unit circle $|z|=1$ as a limit $N \to \infty$.
(See Erdos-Turan type theorem in appendix \ref{app:theorems}.)
The relation between the distribution of zeros of the polynomials
with random coefficients and the pole-zero pairs distribution of
the Pade approximated function is interesting and future problem.

\section{Detection of singularities by expansion-center-shift}
\label{sect:direct-nb}
In this section, we give a simple idea to investigate the singularity
of the functions based on estimation of the
convergence radius of power expansion about the shifted center.
This numerical method can take up the slack result
by the Pade approximation.

\subsection{expansion-center-shift}
We consider the following function $F(z)$ with convergence radius $R_c=1$,
which is expanded about $z=0$ as follows,
\begin{eqnarray}
\label{eqA-1}
F(z) = \sum_{n=0}^\infty a_n z^n.
\end{eqnarray}
Here we would like to investigate whether
the circle $|z|=1$ is a NB or not.
To this end, we shift the expansion center from about origin to around
the $|z|=1$, and estimate the convergence of the
new expansion.
Then, new expansion about the center $\omega=(1-\epsilon) e^{i\phi}$
inside of the convergence domain $|z|<1$
becomes,
\begin{eqnarray}
\label{eqA-2}
F(z) = \sum_{m=0}^\infty b_m (z-\omega)^m,
\end{eqnarray}
where $\epsilon (<1) $ denotes the distance
from the convergence edge $|z|=1$ and $\phi$ is the argument of the center.
We can express the coefficients $\{ b_m \}$ in terms of
the coefficients $\{ a_n \}$ of the truncated Taylor series
Eq.(\ref{eqA-1}) as follows:
\begin{eqnarray}
&& b_m \equiv F^{(m)}(\omega)/m! \simeq \sum_{n=m}^{N} F_{mn},
\label{eqA-4-1}
\end{eqnarray}
where 
\begin{eqnarray}
F^{(m)}(\omega) &=& \frac{d^m}{dz^m}F(\omega), \\
F_{mn} &=& \frac{n!}{m!(n-m)!}\omega^{n-m}a_n.
\end{eqnarray}

The new convergence radius of the function $F(z)$
can be given by the coefficients $\{ b_m \}$ of
the center-shifted expansion as,
\begin{eqnarray}
\label{eqA-3}
R(\omega) &=& \lim_{m \rar \infty} R_\omega(m) \\
 &\equiv &  \lim_{m \rar \infty} \frac{1}{\sqrt[m]{|b_m|} }.
\end{eqnarray}

\begin{figure}[htbp]
\begin{center}
\includegraphics[width=5.5cm]{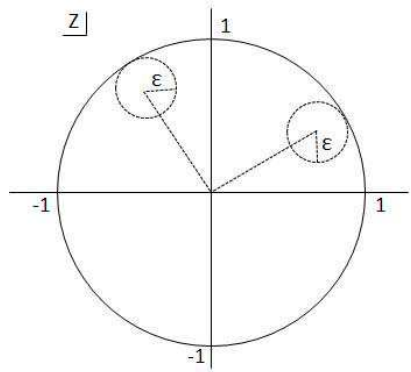}
\caption{
(Color online)
Illustration of expansion-center-shift to confirm
the existence of natural boundary on $|z|=1$.
$\epsilon$ is the distance between the center $\omega$ of the expansion
and the unit circle $|z|=1$.
}
\label{fig:circle}
\end{center}
\end{figure}

Accordingly, if the convergence radius becomes $R(\omega)=\eps$
for any angles $\phi $ of deviation
when the center $\omega$ of the expansion
is shifted close to 
the circle $|z|=1$ from its inside,
we can guess the unit circle $|z|=1$ is
the NB of the function. (See Fig.\ref{fig:circle}.)
We will call this direct method of searching for the singularities
as "expansion-center-shift".

In appendix \ref{app:direct-method}, we give some technical remarks
when we numerically apply the expansion-center-shift.

\subsection{Application of the expansion-center-shift
to functions with poles and NB}
Here, we give some numerical examples to which
the expansion-center-shift is applied.
In the case of lacunary series the nonzero expansion coefficients
become more and more sparse with increase in the order of
expansion, and thus it does not afford significant numerical result
for the expansion-center-shift.

\subsubsection{Example 1: artificially constructed functions with poles}
As a first example
we consider the following artificially constructed functions
with some poles on the circle $|z|=1$ as,
\beq
f_{pole}(z) = \sum_{k} \frac{1}{z-z_k},
\eeq
where $z_k=\exp{i \theta_k}$.

\begin{figure}[htbp]
\begin{center}
\includegraphics[width=7.5cm]{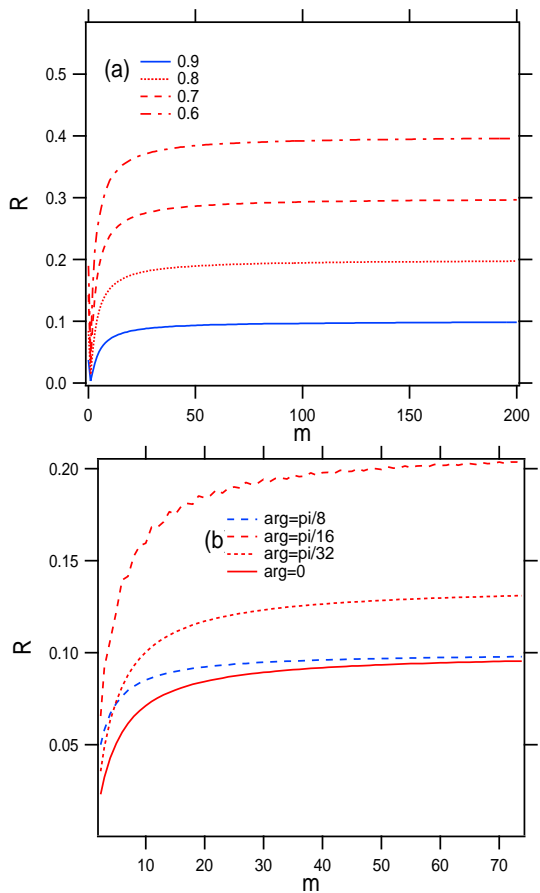}
\caption{
\label{fig:art-pole1}(Color online)
Convergence property of $R(m)$ for an artificially
constructed function $f_{pole}(z)$
with the poles at $z_0=1$, $z_{\pm 1}=\exp \{\pm i \pi/8 \}$.
(a)The cases of the radius $\eps=0.4, 0.3,0.2,0.1$
at a fixed $\phi=0$.
(b)The cases of the argument $\phi=0, \pi/32, \pi/16, \pi/8$
at a fixed $\eps=0.1$.
}
\end{center}
\end{figure}

Figure \ref{fig:art-pole1} shows the $m-$dependence of the convergence
radius given by (\ref{eqA-3}) for $f_{pole}(z)$
with poles at $z_0=1$, $z_{\pm 1}=\exp \{\pm i \pi/8 \}$.
It is found that the convergence radius $R$ smoothly
converges to $\epsilon$ when the location of the center
$\omega$ of expansion is on the real axis, i.e. $\phi=0$,
as seen in Figure \ref{fig:art-pole1}(a).
On the other hand, Figure \ref{fig:art-pole1}(b) shows
the convergence radius $R(m)$ as a function of $m$ 
when the arguments of
the expansion center are taken as $\phi=0, \pi/32, \pi/16, \pi/8$
at a fixed $\eps=0.1$.
As is expected, it asymptotically converge
to the
the distance between the expansion
center $\omega$ and the location of
the nearest pole.
The distances are $0.1,0.137,0.211,0.1$
for $\phi=0, \pi/32, \pi/16, \pi/8$, respectively.

\begin{figure}[htbp]
\begin{center}
\includegraphics[width=6.5cm]{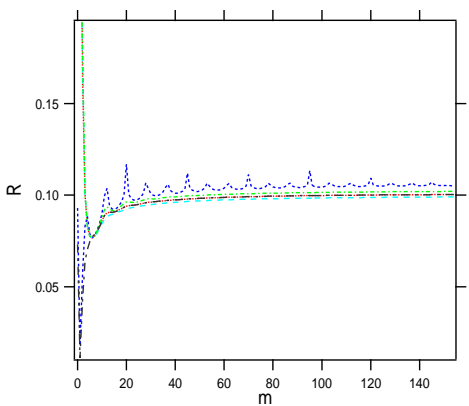}
\caption{
\label{fig:art-pole2}(Color online)
Convergence property of $R(m)$ for the an artificially
constructed function $f_{pole}(z)$
with 80 poles symmetrically distributed
with a regular interval on the unit circle $|z|=1$.
These results are shown for five randomly
chosen arguments $\phi$ within $[0,\pi/4]$ and $\eps=0.1$.
}
\end{center}
\end{figure}

Figure \ref{fig:art-pole2} shows the results for
an artificially constructed function with 80 poles
symmetrically distributed on the unit circle $|z|=1$.
The convergence radius approach the value $\epsilon$ regardless of
the argument $\phi$ of the expansion center.
The fluctuation is cancelled due to superposition of the contribution
from the many poles in the almost all cases.

As a result, the expansion-center-shift can numerically
detect the singularities of the functions with poles on the unit
circle $|z|=1$.

\subsubsection{Example 2: Random power series} 
As a second example,
let us apply the expansion-center-shift for
the random power series which has a NB
on the unit circle $|z|=1$
with probability 1.

\begin{figure}[htbp]
\begin{center}
\includegraphics[width=6.5cm]{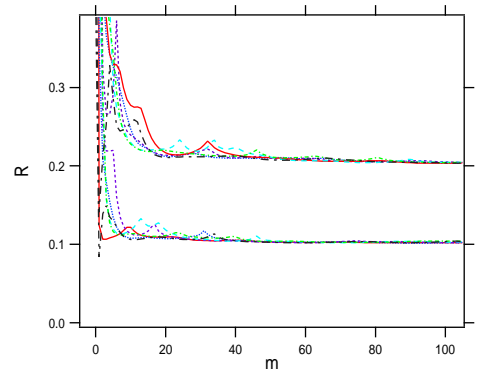}
\caption{
\label{fig:ran-di}(Color online)
Convergence property of $R(m)$ for the random power series.
These results are $N=2000$ and $\epsilon=0.1,0.2$
for 6 different some arguments.
}
\end{center}
\end{figure}

Figure \ref{fig:ran-di} shows the
the variation $R(m)=|b_m|^{-1/m}$(Eq.(\ref{eqA-3})),
at 6 different arguments of the expansion center
taking $\eps=0.2$ and $\eps=0.1$.
$R(m)$ converges to $\eps=1-|w|$ for all angles,
which is consistent with the fact that $|z|=1$ is the NB.

At the end, we list up some noteworthy facts when we use the
expansion-center-shift:
(i) it is applicable to detect singularity on a circle $|z|=1$
by using the rescaled expansion coefficients
$\{ a_n \}$ of order $O(1)$.
(ii)it is a complementary method for Pade approximation
which suggests the location of the singularity by the distribution of
the poles and zeros around on $|z|=1$.

\section{Analyticity of some quantum states I:Pade Approximation}
\label{sect:pade-quantum_nb}
In the following two sections, it will be shown that
NB of the quantum wavefunctions can be
numerically detected by
Pade approximation and expansion-center-shift.
In this section, we investigate the singularity of
wavefunction by Pade approximation.
To our best knowledge, this is the first attempt
to characterize the singularity of
wavefunction in the complex plane.

\subsection{Basic setting}
We investigate the eigenstates
of one-dimensional tight-binding model in the site representation of
Shr$\ddot{o}$dinger equation $H|\Psi>=E|\Psi>$ 
a typical quantum states $\{ u(n) \}$,
\begin{eqnarray}
u(n-1)+u(n+1)+ V_n u(n) =E u(n),
\label{eq:tight-binding}
\end{eqnarray}
where $u(n)=<n|\Psi>$,
and $E$ and $V_n$ are energy of the system and the on-site energy,
respectively.
In this paper, fixed boundary condition, $u(0)=u(N)=0$,
is used in order to make the all the
amplitudes $\{ u(n) \}$ real values, for simplicity.
Here, $N$ denotes the system size.
The following expression by the transfer matrix $T_n$ is sometime
convenient.
\begin{eqnarray}
\left(
\begin{array}{c}
u(n+1) \\
u(n) \\
\end{array}
\right)
= T_n
\left(
\begin{array}{c}
u(n) \\
u(n-1) \\
\end{array}
\right),
\end{eqnarray}
where
\begin{eqnarray}
T_n=
\left(
\begin{array}{cc}
E-V_n & -1 \\
1 & 0 \\
\end{array}
\right).
\end{eqnarray}
The Lyapunov exponent (Landauer exponent) is then defined
in the thermodynamic limit $N \to \infty$ as,
\begin{eqnarray}
\gamma=\lim_{N \to \infty}
\frac{\log \parallel \prod^N_{n=1} T_n \parallel}{2N},
\label{eq:product}
\end{eqnarray}
where $\parallel . \parallel $ is a matrix norm.
The localization length $\xi$ is given by $\xi=1/\gamma$.
This growth is also related to
growth/decay properties of generalized eigenfunctions associated with the
Schrodinger equation.

The Fourier representation of the eigenstate $u(n)$ is given as,
\begin{eqnarray}
\label{prepre}
\Psi(p) = \frac{1}{\sqrt{N}} \sum_{n=1}^{N} e^{-ip n} u(n),
\end{eqnarray}
where $p=2\pi P/N, P=0,1,...,N-1$.
It converges to an analytic function of continuous
real variable $p$ by taking the limit
$N\to\infty$
if the series $u(n)$ has a finite radius of convergence numerically.
(A concrete example of convergence in the limit $N \to \infty$
is demonstrated for Harper model in Fig.\ref{fig:eig1}.)
Moreover, we can straightforwardly extend the real continuous function
 $\Psi(p)$ by taking the limit $N \to \infty$ 
into the complex plane
and shifting $p \to p+i\eta$, 
where $\eta (>0)$ is a real variable.

Before that,
we consider $\Psi(p+i\eta)$
when the eigenfunction $\{ u(n) \}$ is exponentially localized
around a center of the localization $n_0(=0)$
in the lattice space.
\beq
\Psi(p+i\eta) = \frac{1}{\sqrt{N}} \sum_{n=-N_L}^{N_R} \e^{-inp}
\e^{\eta n}u(n),
\label{eq:rescale1}
\eeq
where $N_L$ and $N_R$ denote the left- and right-extremity
of the lattice space, respectively. ($N=N_L+N_R$ and $N_L=N_R$
in our case.)
We introduce a complex variable $z$ given by
\beq
z \equiv \e^{-ip }\e^{\eta } = \e^{-i(p+i\eta) },
\eeq
and divide the function $\Psi(z)$
into two parts as follows,
\beq
\Psi(z) &=& \Psi_+(z)+\Psi_-(z), \\
\Psi_+(z)
&=& \frac{1}{\sqrt{N}} \sum_{n=0}^{N_R} z^n u(n), \\
\Psi_-(z)
&=& \frac{1}{\sqrt{N}} \sum_{n=-1}^{-N_L}z^n u(n).
\eeq
Here, $\Psi_-(z)$ is analytic in p-axis
because the factor $e^{\eta n}$ works as the convergence factor for $n < 0$,
and the singularity of the quantum state $\Psi(z)$ comes from $\Psi_+(z)$.

In general, we have to use the expansion coefficients with the order $O(1)$
to keep the numerical accuracy of Pade approximation as far as possible,
and to make the radius of convergence in order $O(1)$.
Accordingly, we should apply Pade approximation
to $\Psi_+(z)$
by using
the rescaled coefficients $\{u(n)e^{\gamma n} \}$
of the exponentially localized quantum state
and the rescaled
variable $z/e^{\gamma} \to z$
in Eq.(\ref{eq:rescale1})
where $\gamma$ is the Lyapunov exponent
of the localized state.

\subsection{Impurity states} 
\begin{figure}[htbp]
\begin{center}
\includegraphics[width=9.0cm]{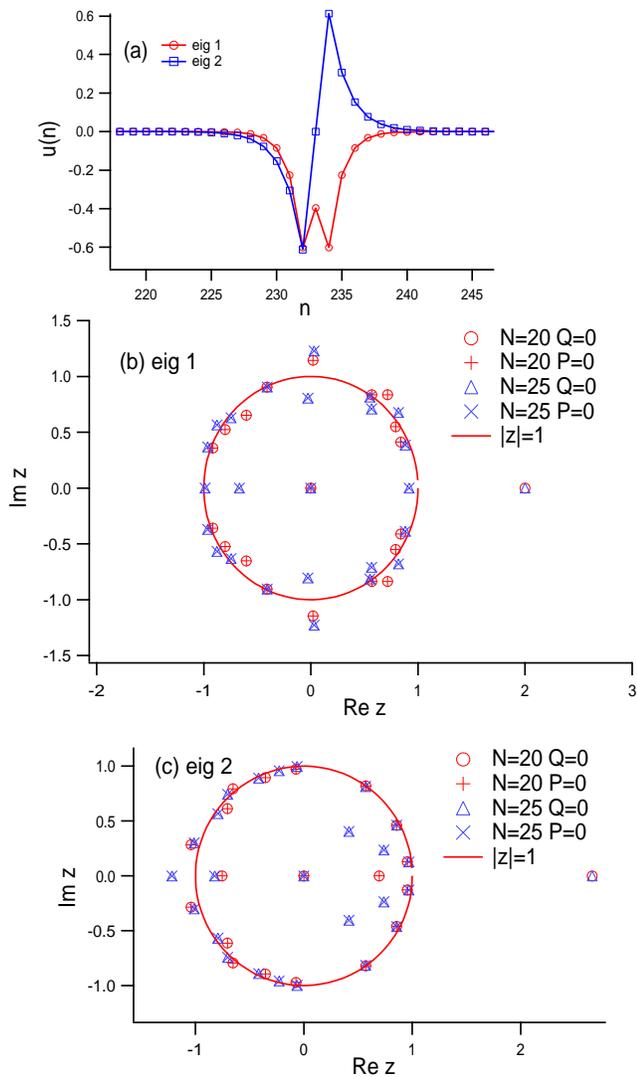}　
\caption{
\label{fig:fig9}(Color online)
Pade approximation for impurity localized states.
(a)Typical symmetric impurity localized states($eig 1$)
and the anti-symmetric one($eig 2$).
(b),(c)Distribution of poles ($\bigcirc, \bigtriangleup$) and
zeros ($+,\times$) of
$[20|20]$ and $[25|25]$ Pade approximations for the two donor type
impurity states.
The unit circle is drawn to guide the eye.
}
\end{center}
\end{figure}

First of all, we examine the impurity localized states
which is obtained
by changing the value of the on-site energy
at only two adjacent sites
to different value from the other sites
in the tight-binding model.
Generally, the impurity problem can be exactly solved and
the localized state decay exponentially
from the localized center at the impurity site
to the outward. (See Fig.\ref{fig:fig9}(a).)
In this case, we used the amplitude $\{u(n) \}$ without the scaling by
the exponential factor
because
it is particularly useful for distinguishing the true pole on real axis
from the ghost ones around $|z|=1$
as shown below.

Figure \ref{fig:fig9}(b) and (c) show
the distribution of the pole-zero pairs
of the diagonal Pade approximated function
for the impurity localized states.
There are poles inside and outside of the unit circle $|z|=1$.
It follows that all zero-pole pairs are canceled with each other
making ghost zero-pole pairs except for an isolated pole
(and an isolated zero) on the real axis.
The location of the true poles are stable for changing the order of
the Pade approximation.
Apparently, the impurity state with exponential localization
can be characterized by a simple pole-type singularity 
in the Pade approximation.
The distance of the isolated true pole from the origin gives
the localization length of the impurity state.


\subsection{Localized states of Harper model} 
In this subsection, we apply Pade approximation
to the localized eigenfunction of Harper model,
in which the sequence of the on-site energy oscillates
as $V_n=2V\cos(2\pi \alpha n )$
in the tight-binding model Eq.(\ref{eq:tight-binding}), 
where $\alpha$ is an irrational number, and $V$ is the potential strength.
Due to the oscillation of on-site potential
incommensurate with the lattice site all of the eigenstates are
exponentially localized for $V>1$, but they are extended for $V<1$.
The localization length is given by $1/\ln |V|(=\gamma^{-1})$ for $V>1$.
Localization means that the eigenfunction is normalizable, ie., 
$\sum_n |u_n|^2<+\infty$.
At the critical parameter value $V=1$, the eigenfunctions exhibit
multifractal and their spectrum is
conjectured to be singular continuous. Further,
a quite interesting fact
is that a diffusive motion very close to a normal
diffusion is realized in the wavepacket propagation.

\begin{figure}[htbp]
\begin{center}
\includegraphics[width=8cm]{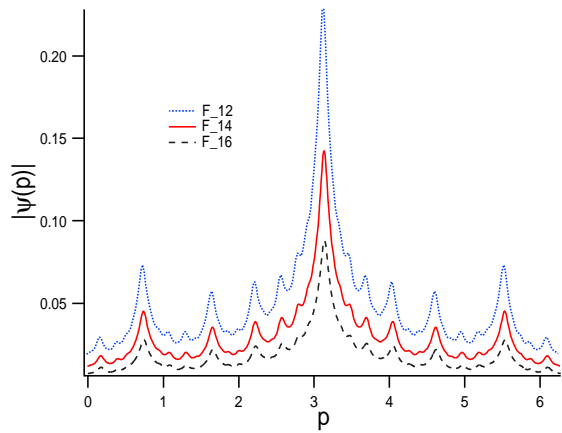} 
\caption{(Color online)
The absolute value of the Fourier transformation $|\Psi(p)|$ of a
localized state ($V=1.05$) of Harper model.
All eigenfunctions are symmetric or antisymmetric 
because we set the system symmetric configuration the in the lattice space. 
The results are shown for $N=2F_{12}$, $N=2F_{14}$, $N=2F_{16}$.
The ground states are used in each case for simplicity.}
\label{fig:eig1}
\end{center}
\end{figure}

In general, eigenstates of the quasiperiodic systems are studied by
using a rational approximation of the irrational number.
Here we used Fibonacci rational approximation for the golden mean
as $\alpha=F_{m-1}/F_{m}$.
In practice, Fig \ref{fig:eig1} shows the discrete
Fourier series expansion $\Psi(p)$ of the
localized eigenstates of Harper model
for three-different system size $N$.
It is found that the Fourier series expansion $\Psi(p)$ of 
the localized state converges a continuous function with 
self-similar fluctuation 
in a limit $m \to \infty$ of the Fibonacci approximation.

For $|z|<1$ the series $\Psi_+(z)$ converges absolutely and the limit
$N \to \infty$ can be taken without any problem.
Accordingly, in the numerical approach for the localized states,
 we can expect that the $\Psi_+(p)$ becomes an analytical function
in a limit $N \to \infty$.

\begin{figure}[htbp]
\begin{center}
\includegraphics[width=8cm]{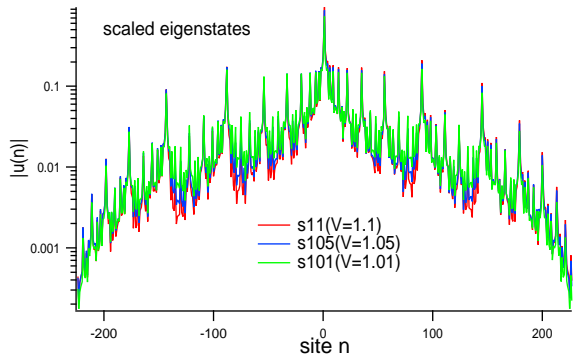}
\caption{
\label{fig:fig11}(Color online)
The eigenstates scaled by the Lyapunov exponent $\gamma$, inverse
localization length, of the Harper model with $V=1.1,1.05,1.01$.
The ground states are used in each case.
}
\end{center}
\end{figure}

In addition, fortunately, in the case of the ground state of Harper model,
the localized state continues to the critical state  
very smoothly as a limit $V \to 1+$ as seen in  Fig.\ref{fig:fig11}. 
Accordingly, we can conjecture that even the localized state
has essentially the same multifractal singular structure as the
critical state if the exponential decaying factor is eliminated,
which means to observe the localized state at the border of the
analytic domain in the complex plane.

This observation motivates us to investigate the analyticity 
of a function $\Psi_+(z)=\sum_n s_n z^n$ 
of the localized eigenstates by using the scaled amplitude
$s(n) = u(n)e^{\gamma |n|}$ for the eigenstate.

Note that, as mentioned in the last subsection, the factor
$e^{-|n|\gamma}$ works as the convergence factor
of the series $\sum_n u_n z^n$, and
the Lyapunov exponent $\gamma$ corresponds to
the depth $\eta$ of the analytic domain
of the $\Psi_+(p)$ in the complex $p-$plane.
As a result, we can directly study complicated fluctuation of the
eigenstates
by the rescaling procedure, which are associated with the singularity
\cite{ketoja95}.

\begin{figure}[htbp]
\begin{center}
\includegraphics[width=8cm]{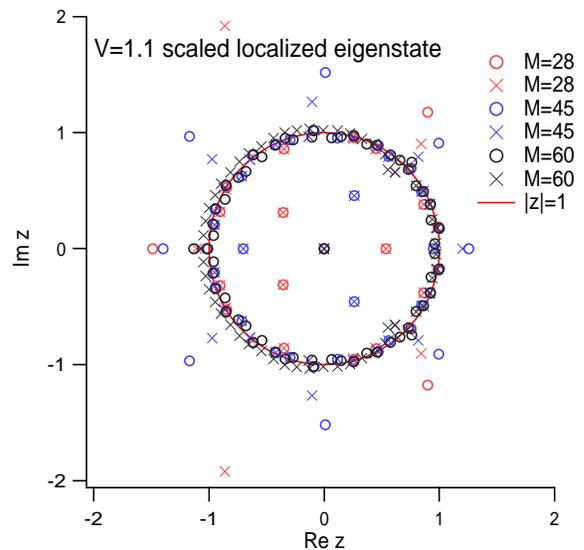}
\caption{(Color online)
Distribution of poles ($\bigcirc$) and zeros ($\times$)
of the $[28|28],[45|45]$ and $[60|60]$ Pade approximations
for the scaled localized state ($V=1.1$).
The result is for the eigenstate in Fig.\ref{fig:fig11}.
The unit circle is drawn to guide the eye.
}
\label{fig:fig13}
\end{center}
\end{figure}

\begin{figure}[htbp]
\begin{center}
\includegraphics[width=7cm]{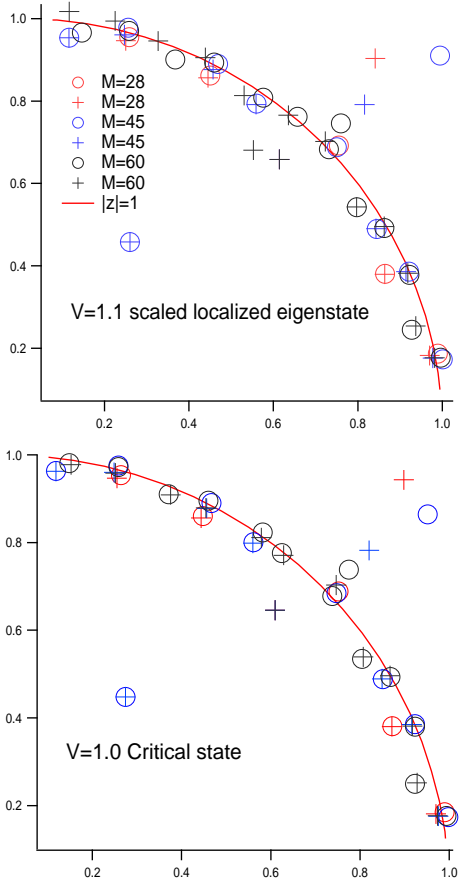}
\caption{(Color online)
(a)An expansion of Fig.\ref{fig:fig13} and
(b)the critical state($V=1.0$) of Harper model.
}
\label{fig:fig13-exp}
\end{center}
\end{figure}

We apply Pade approximation to the localized eigenstate of Harper model,
where the ground state of Harper model with $V=1.1$ is taken as a
typical example.
As are seen in Fig.\ref{fig:fig13} and Fig.\ref{fig:fig13-exp},
the all zero-pole pairs are perfectly cancelled inside
the circle $|z| < 1$ , on the other hand, around $|z|=1$
and outside the circle the pairs are not cancelled.
The distribution of the poles tend to cluster
around $|z|=1$, which strongly suggests
that $|z|=1$ is a NB as $N \to \infty$.

\begin{figure}[htbp]
\begin{center}
\includegraphics[width=8cm]{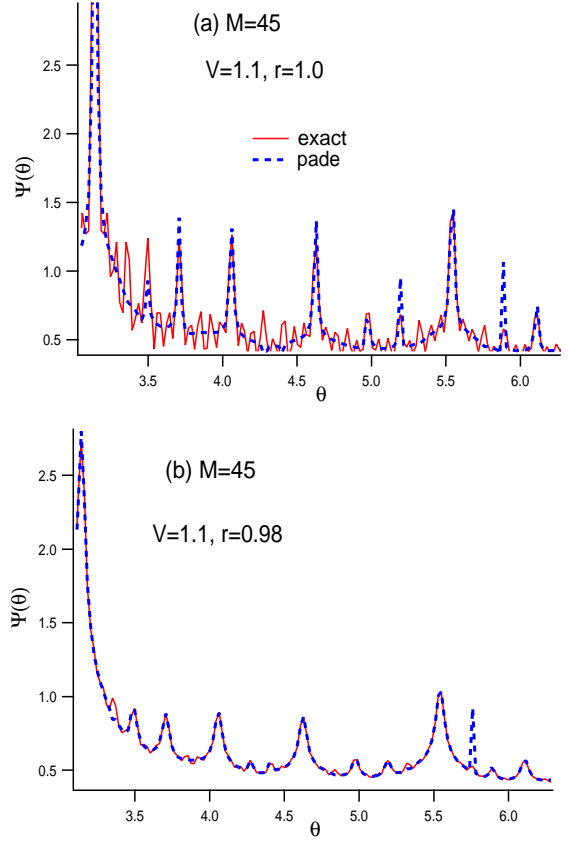}
\caption{
\label{fig:fig14}(Color online)
Comparison of exact and the $[45|45]$ Pade approximated eigenstates
in $\theta-$representation
$\Psi_r(\theta)\equiv \Psi(r\e^{i\theta})$
fixed at (a)$r=|z|=1.0$, (b)$r=|z|=0.98$.
Here the ground state of Harper model with $V=1.1$ is used.
}
\end{center}
\end{figure}

Figure \ref{fig:fig14} shows
Pade approximated eigenfunctions and the exact eigenfunctions
in the $\theta-$representation $\Phi_r(\theta)=\Phi(r\e^{i\theta})$
for the eigenstate of the Harper model with $V=1.1$.
The similar result holds for the critical state
because the shape of the scaled wavefunction
of the localized state is almost similar to the critical state.
The peak structures of the critical state and of
the localized state almost coincide with each other
even on the unit circle $|z|=1$($r=1$), and
they perfectly coincide in the case of the
$r=0.98$, which is
very little inside
the unit circle.
These results strongly suggest that the localized ground state
of Harper model has a NB on $|z|=1$ irrespective
of the value of $V$ when we rescale the variable by the
Lyapunov exponent.
However, we should pay an attention to the following facts
in the Harper model.
(a)All the eigenstates have same localization length
if the potential strength $V$ is same.
(b)In the localization cases ($V>1$)
the eigenstates rescaled by the Lyapunov exponent have
the similar structure as the corresponding critical state ($V=1$),
regardless of the potential strength.
(c)The most important feature is self-dual at the potential strength
$V=1$ (Aubry duality):
the equation for the Fourier transform $\Psi(p)$ of $u(n)$ is identical to
Eq.(\ref{eq:tight-binding}). The quantum eigenstates are extended for
$V< 1$,
exponentially localized for $V > 1$
and at the critical value $V=1$, the states are power-law localized.
Accordingly, all claims for $V>1$ hold for the cases of $V<1$ in
the conjugate representation.

As a result, we can expect that all the eigenstates of Harper model
have a NB on $|z|=1$ in the limit $N \to \infty$
if we rescale by the Lyapunov exponent and transform the eigenfunctions.
The scaling by the Lyapunov exponent
($\gamma \to 1/\log V$) is equivalent to
elimination of the exponential factor from the amplitude
of the eigenfunctions and to access to the convergence of radius
of the series $\Psi(p)$.

Thus the complex structure of wavefunction observed
on the real space only at the critical value $V=1$,
which yields a normal diffusion that is seemingly a
 time-irreversible dynamics.
However, the structure potentially exists in the
complex plane as the NB even in the localized states with $V$ larger
than the critical value.

Finally we would like to
point out that the rescaled localized states $s(n)$
satisfies the following equation,
\begin{eqnarray}
e^{-\gamma}s(n+1)+V_n s(n) + e^{\gamma}s(n-1) = E s(n).
\end{eqnarray}
Ketoja and Satija have directly investigated the fluctuation of
the eigenstate of the scaled Harper equation.
\cite{ketoja95,satija99,Jazaeri01}.
This is nothing more than the eigenstates
in non-Hermitian Harper model \cite{hatano96}.

\subsection{Anderson-localized states} 
The localized states of Harper model, 
as well as the Bloch states and the impurity states,
 are specific
in the sense that the on-site energy follows an
artificially constructed simple sequence.
In this subsection, we deal with Anderson model that
describes a more natural random systems with a statistically
translational invariance.

All the eigenstates are exponentially localized when
the sequence of the on-site energies $\{ V_n \}$ are generated by
i.i.d. random variables in the tight-binding model
in Eq.(\ref{eq:tight-binding}).
We take uniform distribution $V_n \in [-W,W]$, where $W$
denotes the disorder strength.
Furstenberg's convergence theorem on the
product of matrices in Eq.(\ref{eq:product}) ensure
the positive definite value of the Lyapunov exponent:
\beq
Prob(\gamma >0) = 1,
\eeq
with probability 1 for almost every energy
if $\{ V_n \}$ are i.i.d. random variables.

\begin{figure}[htbp]
\begin{center}
\includegraphics[width=9.0cm]{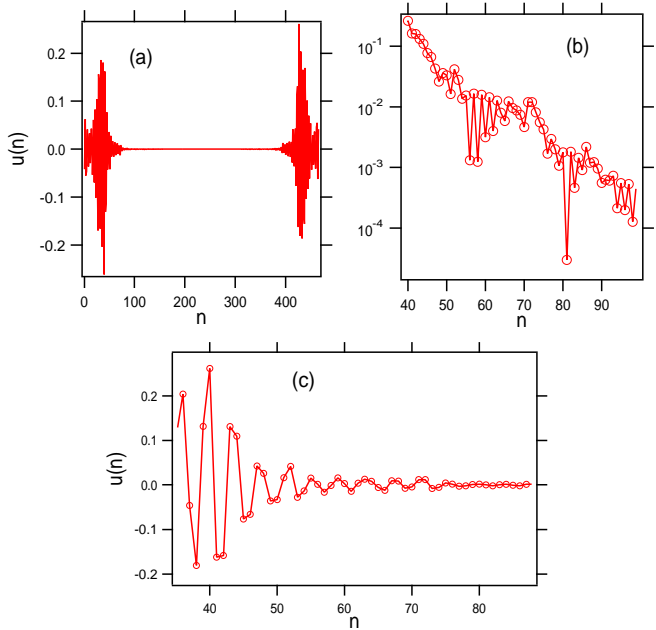}
\caption{
\label{fig:fig10}(Color online)
(a)A typical ${ u(n) }$ of Anderson-localized state
at band edge in the Anderson model with $N=512$ and $W=1$.
(b)A part of the exponentially
localized state $|u(n)|$ in the semi-logarithmic scale.
(c)A part of the amplitude of the localized state used in
Pade approximation.
}
\end{center}
\end{figure}

Let us apply the Pade approximation to a 
typical Anderson-localized state with exponential decay
decorated by complicated fluctuation, as seen in 
Fig.\ref{fig:fig10}(a). 
Note that 
it is a typical double-hump state with two exponential peaks \cite{mott70}.
Here we rescale an exponential decay part of 
the double-hump state with shape of two exponential peaks
in Fig.\ref{fig:fig10}(b) and (c) 
by the numerically estimated Lyapunov exponent, 
and apply the Pade approximation for the rescaled state.

\begin{figure}[htbp]
\begin{center}
\includegraphics[width=8.0cm]{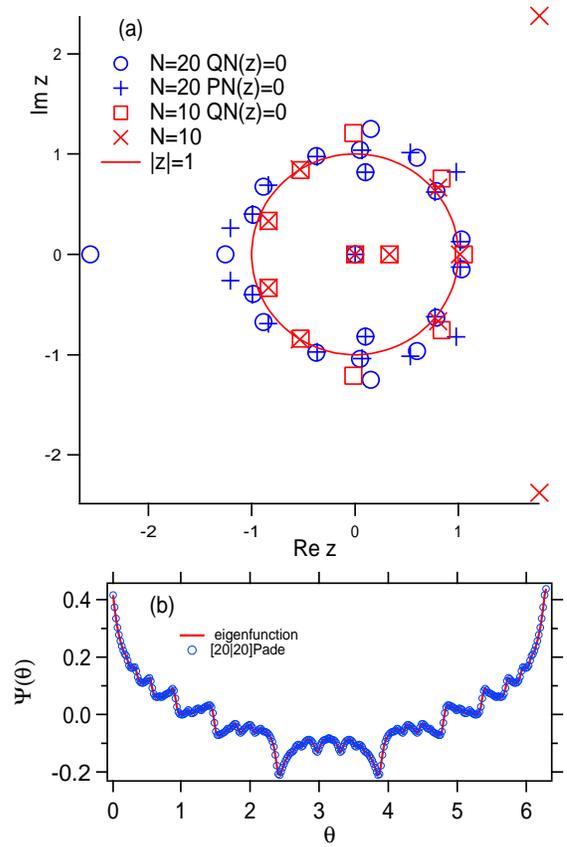}
\caption{(Color online)
\label{fig:fig10-2}
(a)Distribution of poles ($\bigcirc, \Box$) and zeros ($+, \times$) of
the $[10|10]$ and $[20|20]$ Pade approximated functions
for the localized eigenstate.
The unit circle is drawn to guide the eye.
(b)The $[20|20]$ Pade approximated functions
and the original wavefunction
in $\theta-$representation with $r=1.0$.
Note that only a part of the eigenstate used in
the Pade approximation is shown.
}
\end{center}
\end{figure}

Figure \ref{fig:fig10-2}(a) shows
the pole-zero pairs of Pade approximation
for the typical Anderson-localized state.
All pole-zero pairs are cancelled inside the circle $|z|=1$
as is the case of the impurity localized state in Fig.\ref{fig:fig9}.
However, it follows that in the Anderson-localized state
some poles around the unit circle remains, and they
are not canceled with zeros. The distribution of poles and
zeroes seems to suggest the existence of the NB
on $|z|=1$, but we can not give a definite conclusion.
This is because, unlike Harper model, the fluctuation around the
exponential decay of localized wavefunction is so wild
that it mask the systematic decay within the number of sites
available for finite-order Pade approximation.
The suggestion, however, can be strongly supported by the
complementary analysis in Sect. \ref{sect:direct-quantum-nb} and
some considerations in Sect.\ref{sect:quantum-singularity}.

As have been done for the simple examples with a NB in Sect.\ref{sect:pade}
and also for the Harper model in the last subsection,
we confirmed the validity of Pade approximation
by comparing the Pade-approximated eigenfunction
with the original eigenfunction
in the angle representation.
Figure \ref{fig:fig10-2}(b) shows the Pade approximated function
can well reproduce the corresponding part of the original
eigenfunction.

Generally, Pade approximation become less accurate with the decrease
in the order of the approximation. In particular, it is less
reliable when the sign of the coefficient fluctuates very wildly
as is the case of the Anderson-localized states.
Accordingly, it is rather difficult to distinguish the type of the
singularity (finite many simple poles or NB) only by Pade approximation whose
order is very limited. (See Fig.\ref{fig:fig10-2}(b))

Before closing this subsection,
we comment on the fluctuation of the wavefunctions.
Multifractal analysis of the fluctuations
of Anderson-localized eigenstates
in the one-dimensional disordered model
shows multifractality of the rescaled eigenstates
by the exponential factor \cite{pietronero87}.
The scaled eigenstates have multifractal property
in both inside and outside localization length.

It is natural to suppose that such a complicated fluctuation is
due to the NB of the Anderson-localized states in the
$p-$representation.
However, unlike the Harper model, it is difficult to 
give a more accurate proof
by a theoretical technique such as renormalization procedure,  
 as shown in the last subsection.

\section{Analyticity of some quantum states II:
expansion-center-shift}
\label{sect:direct-quantum-nb}
In this section, we apply the expansion-center-shift
to the quantum eigenstates to strengthen our claim
in the last section
that the quantum localized states have a NB.
In this method, much higher-order expansion
coefficients than the Pade approximation
are necessary to get accurate results.

\subsection{Impurity states} 
First of all, we apply the expansion-center-shift
to the impurity state rescaled by the numerically estimated
exponential factor.
Therefore, it is convincible that the impurity state is characterized
by a simple isolated pole on the real axis based on a result
of Pade approximation in the last section.

\begin{figure}[htbp]
\begin{center}
\includegraphics[width=7.5cm]{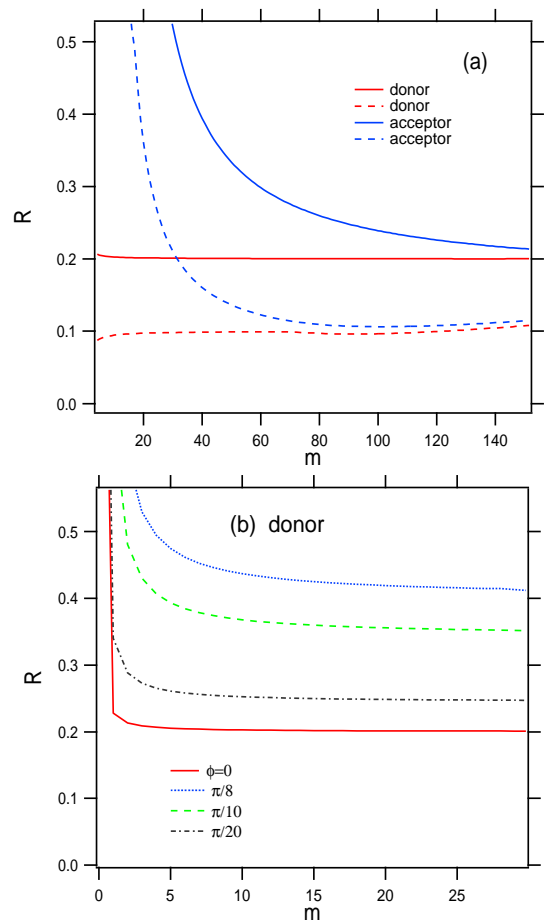}
\caption{
\label{fig:direct-imp}(Color online)
Convergence property of $R(m)$ for the impurity states rescaled by
exponential factor.
(a)The cases of the $\eps=0.2,0.1$ with $\phi=0$ and
(b)the cases of the argument $\phi=0, \pi/32, \pi/16, \pi/8$ with
$\eps=0.1$
are shown.
}
\end{center}
\end{figure}

Figure \ref{fig:direct-imp}(a) shows the results
for the cases of the expansion centers being shifted to
$\omega(=1-\eps)=0.8,~0.9$ on the real axis, i.e. the argument $\phi=0$.
They are shown by solid lines for donor type impurity state
and by dashed lines for accepter type impurity state, respectively.
It is found that in the case of acceptor type impurity state
the convergence of $R(m)=1/\sqrt[m]{|b_m|}$ is 
slower than that of donor type impurity state
due to the alternating sign of the amplitude.
Note that the acceptor impurity state is oscillatory
localized with the same localization length as the donor
impurity state.
In all cases, the radius $R(m)$ converge to the
expected value as $m$ increases.
A convergence problem as seen in Sect.\ref{sect:direct-nb} does not occur
in this case.

Next, fixing $1-|w|=\eps=0.1$, we examined the convergence
property by changing the argument $\phi$ (arg$w$)
for the donor type impurity state.
As shown in Fig.\ref{fig:direct-imp}(b),
$R$ converges well to the
expected value in each case
based on assumption that a simple pole is located at $z=1$.
(Note that we used the rescaled impurity eigenstate.)
In this case the convergence problem for larger $m$
comes up with increase value of the argument $\phi$,
as we discussed in Sect.\ref{sect:direct-nb} and appendix
\ref{app:direct-method}.

As a result, the expansion-center-shift suggests that
the impurity states can be characterized by simple pole at
real axis that is consistent with the result by Pade approximation
in the last section.

\subsection{Localized eigenstates of Harper model} 
We will reexamine whether the unit circle $|z|=1$ is a
NB of the localized state
of Harper model ($V=1.02$) by the expansion-center-shift.
We use the rescaled expansion coefficient $s(n)=\exp(\gamma |n-n_0|)u(n)$
as in the Pade approximation, where
$\gamma(=1/\log V)$ is the Lyapunov exponent and $n_0$ is a 
localization center of the localized state.

\begin{figure}[htbp]
\begin{center}
\includegraphics[width=6.5cm]{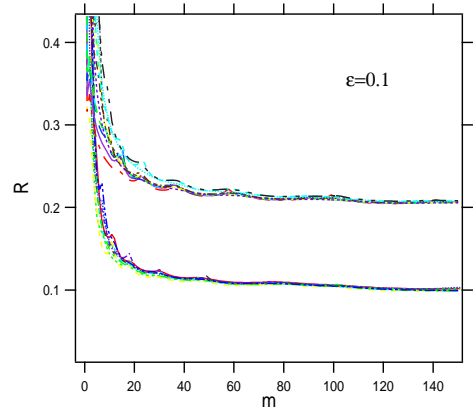}
\caption{(Color online)
\label{fig:multi-exp-harper}
Convergence property of $R(m)$ for the scaled localized state of
Harper model ($V=1.02$).
These results are shown for $N=2F_{16}=1974$ and $\epsilon=0.1, 0.2$
with randomly selected 10 arguments of the expansion center.
}
\end{center}
\end{figure}

The convergence property is shown in Fig.\ref{fig:multi-exp-harper}
by taking $|w|-1=\eps=0.2$ and $\eps=0.1$ for various values
of $\phi={\rm} arg w$.
 For all values of $\phi$ the $R(m)$
 converges to the expected values, namely, $R(m) \to 0.2$ for $\eps=0.2$
and $R(m) \to 0.1$ for $\eps=0.1$, respectively. This fact means that
at any point on $|w|=1-\eps$ the power series expansion of
the wavefunction has the convergence radius $\eps$, and 
the singularities are densely distributed on $|z|=1$.
The result suggests the unit circle is a NB of the
localized eigenstates of Harper model.

\subsection{Anderson-Localized eigenstates} 
\begin{figure}[htbp]
\begin{center}
\includegraphics[width=7.0cm]{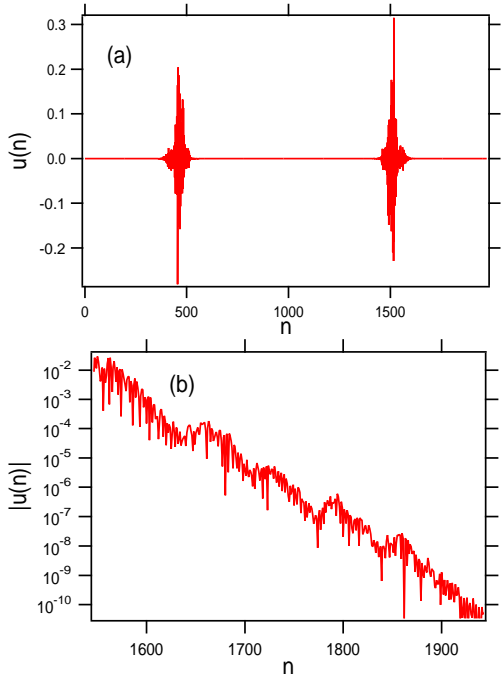}
\caption{(Color online)
\label{fig:and-loc}
(a)A typical Anderson-localized state $\{ u(n) \}$ around the band center
in the Anderson model with $N=2000$ and $W=1$.
(b)A part of the exponentially
localized state $|u(n)|$ in the semi-logarithmic scale
used in the expansion-center-shift.
}
\end{center}
\end{figure}

\begin{figure}[htbp]
\begin{center}
\includegraphics[width=7.0cm]{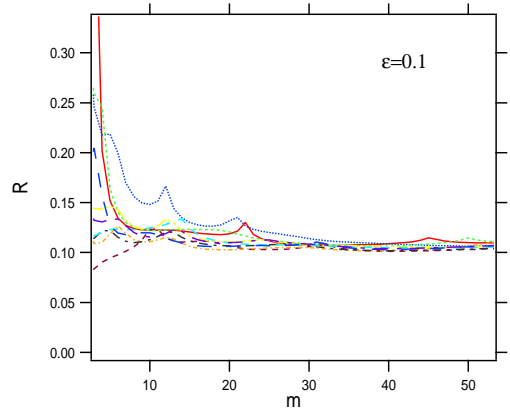}
\caption{(Color online)
\label{fig:multi-exp-and}
Convergence property of $R(m)$ for the rescaled Anderson-localized state in
Fig.\ref{fig:and-loc}.
These results are shown for $W=1$, $N=2000$ and $\epsilon=0.1$
for the expansion-center-shift
with randomly selected 10 arguments of the expansion center.
}
\end{center}
\end{figure}

Figure \ref{fig:and-loc} shows a typical Anderson-localized state
around the band center. The exponentially decaying part used in
the expansion-center-shift is also shown.
We take the rescaled amplitude $s(n)=\exp(\gamma |n-n_0|)u(n)$
of the localized states, where $\gamma$ is the numerically
estimated Lyapunov exponent of the localized eigenstate.
As mentioned in Sect.\ref{sect:direct-nb},
we estimate the convergence radius $R(m)$
of the center-shifted power series about $w=(1-\eps)\e^{i\theta}$.
The convergence property is shown in Fig.\ref{fig:multi-exp-and}
for $\eps=0.1$
and many different arguments of the expansion center.
The convergence of radius of new expansion about the center $\omega$
vicinity of $|z|=1$ strongly suggests the unit circle is a NB of the
rescaled eigenstates as is conjectured by the Pade approximation.

\section{Localized quantum states and NB of the potential}
\label{sect:quantum-singularity}
In this section, we discuss about an interesting relation between
analyticity of the potential $V_n$ of
one-dimensional tight-binding model and the eigenfunctions based on
recently proved mathematical theorems. By the analyticity of potential
$V_n$
we means the analyticity of the $Z-$transform $V(z)=\sum_n V_n z^n$,
namely, the momentum representation of the potential function
if we take $z=\e^{-ip}$. We refer to $V(z)$ as the potential
generating function (PGF) in the present paper.

G-M-P theorem \cite{goldsheid77}
states that one-dimensional Schrodinger equation with an ergodic
and stationary random potential have
a positive Lyapunov exponent ($\gamma>0$)
of the wavefunction with probability 1.
The positive Lyapunov exponent is necessary-sufficient condition
for a pure point set spectrum of the operator
and all the eigenfunctions decay exponentially at $n \pm \infty$.
For the random $V_n$ its PGF has a NB according to the Steinhaus's theorem,
and thus the presence of the pure point spectrum of random Schrodinger
equation corresponds to the existence of the NB of PGF the random series.
Recently, Breuer and Simon discovered a more general correspondence
between the analyticity of PGF and
the spectral property of the Schrodinger operators \cite{breuer11}.

{\bf Theorem 1a (Breuer-Simon 2011):}
Let $a_n(\omega)$ be a stationary,
ergodic, bounded, nondeterministic process. Then for a.e.
$\omega$, $f(z)=\sum_{n=0}^{\infty}a_nz^n$
has a strong NB.
■

On the other hand, in the localization problem of one-dimensional
continuous/discrete Shrodinger operators, Kotani theory and
its extended versions \cite{kotani82,simon83,kotani86} show
more powerful results than G-M-P theorem.

{\bf Theorem 1b (Kotani 1982):}
If the potential sequence $V_n$ is nondeterministic under the conditions
(i)stationarity,
(ii)ergodicity,
(iii)integrability,
then
there is no absolutely continuous (a.c.) spectrum of the operator.
■

The stationarity means a translation invariance of the sequence,
and integrability is the boundedness of the ensemble average.
And the stationarity and ergodicity ensure Kolmogorov $0-1$ law.
In the Kotani's meaning, roughly speaking,
the {\bf non-deterministic} potential sequence means
that left-half part of the sequence can not determine
whole of the right-half part of the sequence a.e. with respect
to the probability measure.

Note that these conditions for the potential sequence is much weaker than
the stationary random potential condition of the localization.
Indeed, it will be the weakest sufficient condition for the occurrence of
the weak localization (weak localization means there is no a.c.
spectrum), and it is quite interesting that these conditions
for the weak localization supports the presence of the NB of the PGF.

The above pair of theorems suggests that there is an interesting
parallelism between the two properties that the potential has a
NB and that the potential yields the weak localization.

An another pair of theorems implying this parallelism is for
an apparently deterministic incommensurate potential
with discontinuities,

{\bf Theorem 2a (Damanik-Killip 2005)\cite{damanik04}:}
Consider the tight-binding model with $V_n(\theta )=g(\alpha n +\theta )$,
where $\alpha$ is irrational, g is bounded and periodic with period 1,
and $g(z) \in [0,1]$.
$g(z)$ is continuous except at finitely many points, at
one of which it has different right and left limit. Then for
a.e. $\theta$, the tight-binding model has no a.c. spectrum.
■

{\bf Theorem 2b (Breuer-Simon 2011):}
Let a function $g(z)$ be as in the Damanik-Killip.
Then for all $\theta $, $f(z)=\sum_{n=0}^{\infty}g(\alpha n +\theta )z^n$
has a strong NB on $|z|=1$.
■

If we agree our assertion that the localized eigenstate has a NB is
generally correct, then the parallelism holds for the two properties
that the PGF has a NB and that the eigenfunction has a NB.

Note that the parallelism is only hold for the sufficient condition
yielding the existence of NB and the occurrence of the weak localization.
However, we emphasize that it is not exactly correct
that whether the PGF has a NB or not  respectively
correspond to the localization or non-localization.
Indeed, the Harper equation has a deterministic, analytic
and incommensurate potential, and its PGF has no NB,
but the Harper equation has localized eigenstates
if the potential strength is large enough.
Furthermore, there is another example.
Suppose that the potential series is a lacunary series,
then its PGF has a NB, but the corresponding sparse potential
sequence $V_n$ consists of
$\{0,1\}$, where the "0" corresponds to ``gap'' of the series,
and the Lyapunov exponent $\gamma$
of the Schrodinger operator 
become zero because the mean gap length
becomes infinite in the limit $N \to \infty$.
As a result, the spectrum is absolutely continuous
and the eigenstates are extended, although the PGF has a NB.
(Note that the result is not contradict with Kotani theory because
the lacunary system is deterministic and nonstationary.)

In spite of the counter examples given above, the parallelism
discovered by Breuer and Simon combined with our result for
localized eigenstates seems to suggest that the presence of the NB 
in PGF and the presence of the NB in eigenstate are strongly 
correlated. 
It may be conjectured that if the PGF has
a NB then at least its eigenfunction also has a NB.

\section{Summary and Discussion}
\label{sect:summary}
To investigate the very origin of complex quantum dynamics,
we studied the analytical property of wavefunction in the quantum
system by complexifying the argument of the eigenstates,
taking one-dimensional tight-binding models, whose eigenstates exhibit
exponential localization as an example.

To capture the singularities of the quantum states
we investigated usefulness of the Pade approximation
and expansion-center-shift for some known functions
which have been proven to have a natural boundary (NB).
The poles and zeros of Pade approximation
accumulate around the NB
with increase in order of the Pade approximation.
The expansion-center-shift compensates for Pade approximation
by detecting the change of convergence radius with shifting
the center of power series expansion.

We used Pade approximation \cite{baker75,baker96}
as a numerical method to investigate
the properties of singularities of Fourier series
(or the Z-transformation)
of the localized and critical states of Harper model
in comparison with the other quantum states such as impurity state
and Anderson localized state.
The impurity states can be characterized by simple poles of
the Pade approximation.
It is found that in the localized states of Anderson model and Harper model
the poles and zeros of the Pade approximated function tend to cluster, forming a NB. 
Next, we confirmed the conclusions
obtained by the Pade approximation are consistent with the result of
expansion-center-shift.

Here, we discuss our original interest on quantum dynamics.
It is by no means a trivial problem whether the same analytical
properties as the eigenfunctions can be observed also for
the time-dependent wavefunction.
If the initial wavefunction is an entire function like Gaussian
wavepacket, then the time-evolved wavefunction is also an entire-function
for a finite time evolution.
We know, on the other hand, empirically that the wavepacket
is also exponentially localized in a limit of long time-evolution,
and the exponential region spreads in time.
As long as we focus on the exponential-decay region, as is shown
in appendix \ref{app:dynamical}, the wavefunction practically looks as if
it has a NB in a similar way to the localized eigenfunctions.

{\it What is physical meaning of the NB of
the localized eigenstates and the dynamically localized states
in Anderson and Harper models?}

We consider the problem based on the localized state of the
Harper model. Let us consider the analytic continued eigenfunction $\Psi(z)$
of Harper model, where $z=\e^{-i(p+i\eta)}$ is a complex variable.
The essential structure of $\Psi(z)$ on the NB is independent
of the potential strength $V$, and thus it does not depend on the
localization length, and it is the same as the critical state
if we rescale as $z\e^{-\gamma} \to z$ by the Lyapunov exponent $\gamma$.

Therefore, we can expect that at the NB the same phenomenon as
the critical state takes place. As $V$ deceases and approaches to
the critical state $V=1$, the NB goes down to the real axis,
and the real axis itself become a NB. The diffusive behavior occurring
in the complex space can be observed in the real world.
Inversely speaking, it may be claimed that a diffusive behavior
of the wavepacket would be realized
already in the complex plane even for $V>1$, in other words,
the localized state is a pre-diffusive state
in the sense that it actually exhibits a diffusive motion
in the complex plane.

With the above scenario in mind we try to go one-step further.
Then we can expect that the similar scenario 
would hold true with the
Anderson-localized states in disordered systems.
Actually, Anderson-localized states have a NB and
the fluctuation of the states show multifractal structure
when the exponential factor is removed by
the scaling of the localization length.
Accordingly, the multifractal structure of the wavefunction
is strongly related to the NB.
The complex structure of eigenstates can be generated by
an accumulation of
the interference effect of the multi-scattering due to the randomness
of the potential.
Therefore, the localization-delocalization transition
observed in higher-dimensional disordered systems and multiple
degrees of freedoms systems can be also interpreted as the result
that NB of the wavefunction approaches the real axis; 
it is the onset of delocalization finally leads
to a normal diffusion and irreversibility of the
time-reversal property. (Note that, unlike the Harper model,
the delocalization do not necessary mean the normal diffusion
in the case of Anderson localization.)

Therefore, we guess that the complex wavefunction
associated with the emergence of the NB might be an ultimate
origin of irreversibility in quantum systems.

As we considered in Sect.\ref{sect:quantum-singularity},
quantum states of spatially extended ``natural systems''
with stationary, ergodic and bounded potential,  
 except for very rare example such as periodic systems,
have a NB, and so it potentially exhibits the onset of 
irreversible phenomenon such as diffusion-like motion 
at the NB. Under certain conditions
the NB, which is usually located in the complex domain, 
falls down on the real space, leading to the onset of 
irreversible phenomenon in the real world. 
Conversely, even though no time-irreversible phenomena is observed 
in the real space, the time-irreversibility may be observed 
in the complex space by ``complexifying'' 
the real coordinate which is the eigenvalue of an observable.
To clarify whether our scenario described
above is true or not we, of course, need an actual proof, 
which is remained as a future problem.


\appendix

\section{Pade approximation of Jacobi lacunary series}
\label{app:lacunary}
In this appendix, we investigate effectiveness of Pade approximation
for a lacunary series $f_{Jac}(z)=\sum_{n=0}^{\infty} z^{2^n}$
with a NB on $|z|=1$, which is called
Jacobi lacunary series after
Jacobi.
Some theorems for
lacunary series with a NB
are given in appendix \ref{app:theorems}.

The Pade approximated function exactly has the following form,
\beq
f^{[2^N]}_{Jac}(z) & \sim & f_{Jac}^{[2^{N-1}|2^{N-1}]}(z) \\
& = & \frac{A^{N}_{Jac}(z)}{1+\sum_{k=0}^{N-2} z^{2^{k}} - z^{2^{N-1}}},
\label{eq:jac}
\eeq
where explicit form of the numerator $A^{N}_{Jac}(z)$
is given as
\beq
A^N_{Jac}(z) &=& z+2z^2 \\ \nn
&+& 2\sum_{n=2}^{N-1}z^{H_n} (z+z^2+\sum_{k=1}^{n-2}
z^{H_{k+2}}),
\eeq
where $H_n=2^{n-1}$.

\begin{figure}[htbp]
\begin{center}
\includegraphics[width=8cm]{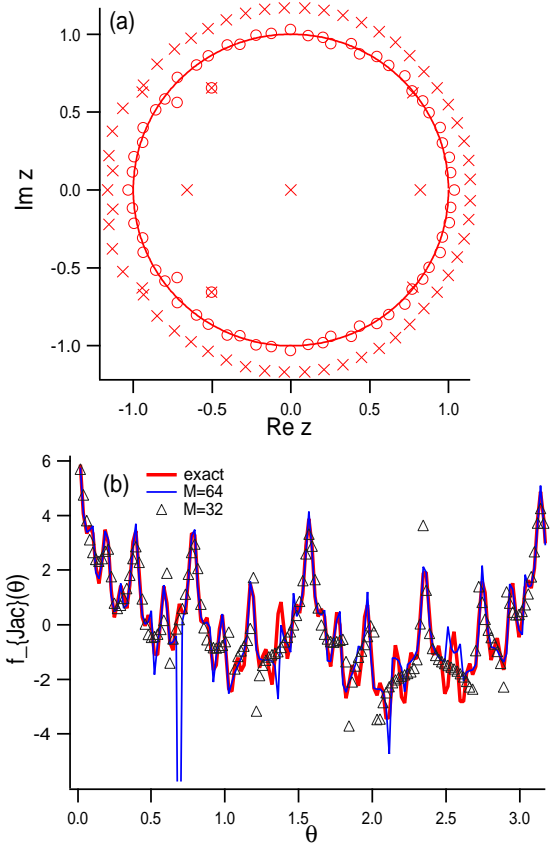}
\caption{
\label{fig:fig3}(Color online)
(a)Distribution of poles($\bigcirc$) and zeros($\times$)
of the $[64|64]$ Pade approximation
for test function $f_{Jac}(z)$ with NB on $|z|=1$.
The unit circle is drawn to guide the eye.
(b)The Pade approximated functions $f_{Jac,r}^{[32|32]}(\theta)$,
$f_{Jac,r}^{[64|64]}(\theta)$ and exact function $f_{Jac,r}(\theta)$
in the $\theta-$representation with $r=1.0$.
}
\end{center}
\end{figure}

Accordingly, the poles of the $[2^{N-1}|2^{N-1}]$ Pade approximated function
are
given by roots of the polynomial,
\beq
1+\sum_{k=0}^{N-2} z^{2^{k}} - z^{2^{N-1}} =0.
\eeq
This is just a lacunary polynomial and the zeros are distributed uniformly
on the unit circle as $N \to \infty$.
In Fig.\ref{fig:fig3} the numerical result of
Pade approximation for $f_{Jac,r}(\theta)$ is shown.
The poles and zeros are plotted for the $[64|64]$
Pade approximation in Fig.\ref{fig:fig3}(a).
In the case of $M=64$ the poles and zeros accumulate around $|z|=1$
as increase of order of the Pade approximation.
Inside the circle $|z|=1$ some cancellations of the ghost pairs appear.
Figure \ref{fig:fig3}(b) shows the Pade approximated functions
in the $\theta-$representation.
It well approximate the original one when the order of Pade approximation
increases.

\section{Some mathematical theorems}
\label{app:theorems}
The following theorems are well-known for function 
with a NB on $|z|=1$.
For the details and proofs,
see the references \cite{korner93,remmert10}.

{\bf Erdos-Turan Type theorem:}
Let us define a random polynomial
\beq
f(z)=\sum_{n=0}^N a_n z^{n},
\eeq
where coefficients $a_n$ are randomly distributed and $a_0a_N \neq 0$.
Then the zeros of the random polynomial cluster uniformly around the
unit circle $|z|=1$ if "size of the truncated series" $L_N(f)$ is
small compared to the order $N$ of the polynomial,
where
\beq
L_N(f) = \log \left( \frac{\sum_{n=0}^{N} |a_n|}{\sqrt{|a_0a_N|} } \right).
\eeq
■

Note that this theorem is also hold for the polynomials with deterministic
coefficients $a_n$ such as Newman type polynomial having coefficients in
the
sets $\{ 0,1 \}$ or $\{ 0, \pm1 \}$.

{\bf Fabry's gap theorem(1899):}
Power series
\beq
f(z)=\sum_{\nu=0}^\infty a_\nu z^{\lambda_\nu}
\eeq
with radius of convergence $R=1$
has a NB on $|z|=1$, provided it is
Fabry series, i.e.
\beq
\lim_{\nu \to \infty} \frac{\lambda_\nu }{\nu} = \infty.
\eeq
■

{\bf Steinhaus's theorem(1929):}
Suppose that the power series $f(z)=\sum_{n=0}^{\infty} a_n z^n$,
has radius of convergence $R=1$.
Let $X_0, X_1,..., X_n$ be a sequence of stochastic variables
obeying i.i.d. in the interval $X_i \in [0,1]$.
Then, with probability one, the power series
\beq
f_{Steinhaus}(z) = \sum_{n=0}^{\infty} a_n w_n z^n,
\eeq
has a NB on $|z|=1$, where $w_k=e^{i 2\pi X_k}$.
■

{\bf Paley-Zygmund's theorem(1932):}
Suppose that $r_0, r_1,..., r_n,...$
be a sequence of stochastic variables taking the binary value
 $r_i = \pm 1$ with equal probability , then the power series
\beq
f_{P-Z}(z) = \sum_{n=0}^{\infty} r_n z^n,
\eeq
has a NB on $|z|=1$.
■

The similar theorem can hold for random power series $\sum_{n=0}^{\infty} r_n z^n$
with a sequence of stochastic variables
obeying i.i.d. in the interval $r_i = [-1,1]$ and $r_i = [0,1]$.

{\bf Szego's theorem (1922):}
Suppose that the power series $f(z)=\sum_{n=0}^{\infty} a_n z^n$,
has radius of convergence $R=1$ 
the values of $\{a_n\}$ is in a finite set, 
then either $|z| = 1$ is a NB,
or else ${a_n}$ is eventually periodic, in which case $f(z)$
is a rational function with poles on $|z| = 1$.
■

Notice that the condition of the potential sequence
in this theorem corresponds to
that in Bernoulli-Anderson model,
which models alloys composed of at
least two distinct types of atoms
by means of the distribution of a Bernoulli random variable
\cite{carmona87,bourgain05}.


\section{Some technical remarks on
the expansion-center-shift}
\label{app:direct-method}
We can calculate the coefficients $\{ b_m \}$ by the coefficients $\{
a_n \}$
of the truncated Taylor series Eq.(\ref{eqA-1}) as follows:
\begin{eqnarray}
\nonumber
&& b_m \equiv F^{(m)}(\omega)/m! \simeq \sum_{n=m}^{N} F_{mn},
\label{eqA-4}
\end{eqnarray}
where
\beq
F_{mn}=\frac{n!}{m!(n-m)!}\omega^{n-m}a_n.
\eeq
We remark some numerical problem in the evaluation of the sum
(\ref{eqA-4}).
By using Stirling's formula, the $n-$dependence of $F_{mn}$ is
represented by
\beq
F_{mn} \propto \frac{n^n}{(n-m)^{n-m}} \omega^n a_n.
\eeq
The modulus $|F_{mn}|$ is well approximated by a
Gaussian function around $n=n_0$
maximizing $n^n/(n-m)^{n-m}|\omega|^n$, which gives
\beq
n_0=\frac{m}{(1-|\omega|)}=\frac{m}{\eps},
\eeq
and the Gaussian approximation becomes
\beq
F_{mn} \sim \exp\{-\frac{\eps^2}{2m}(n-n_0)^2\} (1-\eps)^n \e^{in\phi} a_n.
\eeq
Thus we have to calculate a large number of coefficients $a_n$
such that $0\leq n \leq N$, and $n_0(=m/\eps) \gg N$ with increase in $m$
and/or decrease in $\eps$. 
In addition, even though we may prepare $a_n$ up to
$N \gg n_0$, the second serious numerical problem comes
from the oscillatory nature of $\omega^n=(1-\eps)^n\e^{in\phi}$.

Suppose the simplest and worst case $a_n=1$,
namely $f(z)=\sum_na_nz^n$ has a single pole at $z=1$, then cancellation
due to the sinusoidal oscillation
of $\omega^n$ makes the result
of summation Eq.(\ref{eqA-4})
extremely small as
\beq
\sum_{n=m}^{N} F_{mn} \sim \e^{-\frac{\phi^2m}{2\eps^2}},
\eeq
except for a case $\phi=0$,
which may be incomputable by numerical summation of Eq.(\ref{eqA-4})
if $\eps$ is small and/or $m$ is large.
This is the worst case, and for more general cases where $a_n$ has no
particular regularity, things will be much relaxed.
However, a particular care must be taken for the numerical accuracy
of the summation.

\section{NB of dynamically localized wavepacket}
\label{app:dynamical}
\begin{figure}[htbp]
\begin{center}
\includegraphics[width=8cm]{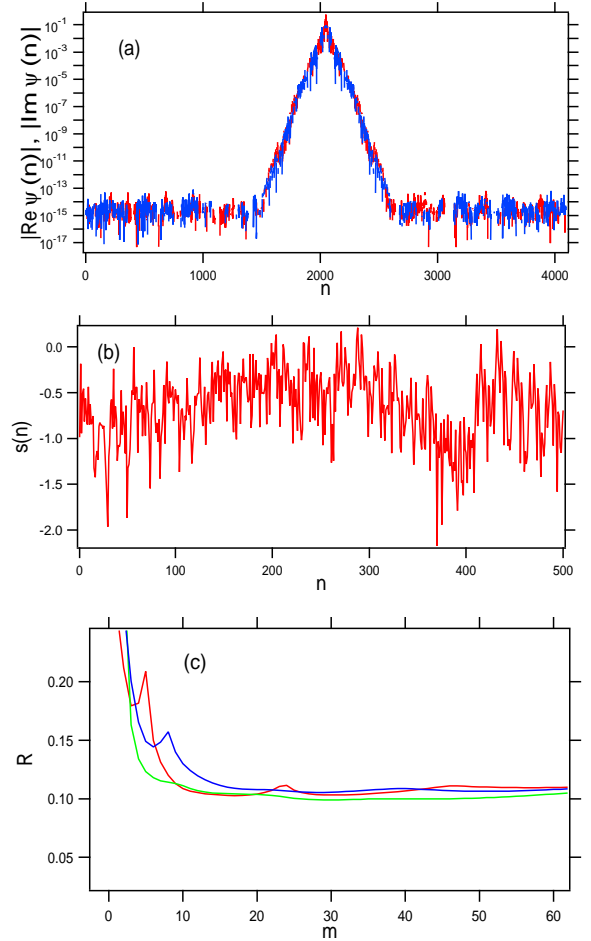}
\caption{
An exponentially localized wave packet of
Harper model with $V=1.04$.
(a)$|Re \Psi(t,n)|$ and $|Im \Psi(t,n)|$ of the wavepacket at $t=200$.
(b)The rescaled amplitude $Re[s(t,n)]$ of the real part $Re[\Psi(t,n)]$
of 1000 sites from the localization center $n_0=2048$.
(c)Convergence property of $R(m)$ for the rescaled localized state
$Re[s(t,n)]$ of
the Harper model for different three arguments.
These results are shown for the system size $N=4096$ and
parameters $\hbar=1$ and $\epsilon=0.1$.
}
\label{fig:haloc-direct1}
\end{center}
\end{figure}

We can expect that dynamically localized states are also
have similar singularity as the localized eigenstates in Harper model
and Anderson model.
Therefore, by the expansion-center-shift, 
we confirm the idea for
the dynamically localized wavepacket in Harper model
numerically obtained by,
\beq
i\hbar \frac{\pr \Psi(n,t)}{\pr t}= \Psi(n+1,t)+ \Psi(n-1,t) \nn \\
+ 2V\cos (2 \pi \alpha n) \Psi(n,t), n=1,2,...,N.
\eeq
with initial state $\Psi(n,t=0)=\delta_{n,n_0}$.
In the Harper model
with the strength $V>1$, spread of the initially localized wavepacket
is suppressed and exponentially localized by the time-evolution.
Accordingly we can obtain the stable localized wavepacket after time fully elapses.
(See Fig.\ref{fig:haloc-direct1}(a).)
Here, we can factorize the exponential decay factor $\e^{-\gamma |n-n_0|}$
of the localized state as 
\beq
s(n,t) = \e^{\gamma |n-n_0|}\Psi(n,t),
\eeq
where we can get the Lyapunov exponent $\gamma$ by numerical fitting for
the dynamically localized quantum states.

We investigate the singularity of the following power series
\beq
F(z)= \sum_n Re[s(n,t)] z^n, 
\eeq
along the unit circle $|z|=1$ by means of the expansion-center-shift,
as in the main text.
The expansion is constructed by the real part of the rescaled amplitude $s(n,t)$
with the order $O(1)$ of the fluctuation.(See Fig.\ref{fig:haloc-direct1}(b).)
We can expect that the series
$\sum_n Re[s(n,t)] z^n$ has a NB on $|z|=1$
due to the fluctuation as in the case of the eigenstates.

Figure \ref{fig:haloc-direct1} shows the result for
the dynamically localized state of the Harper model with $V=1.04$.
The convergence of $R(m)$ for $\eps=0.1$
suggests the unit circle is a NB of the rescaled dynamically
localized state.
Generally, the nature of the fluctuation of the rescaled dynamically
localized
states $s(n)$ is not depend on the potential strength $V$, 
as seen in the cases of the localized eigenstates.
Accordingly we can suggest that the dynamically localized quantum states 
of the Harper model have a NB in the complex plane.

\section*{Acknowledgments}

This work is supported by Kakenhi 24340094 based on the tax of
Japanese people and the authors would like to acknowledge them.
They are also very grateful to Shohji Tsuji and Koike memorial
house for using the facilities during this study.


\end{document}